\documentclass[aps,prd,nofootinbib,showpacs,preprintnumbers,longbibliography,amssymb,11pt]{revtex4-1} 
\usepackage[sort&compress]{natbib}
\usepackage{slashed}

\usepackage{graphicx}
\usepackage{epsfig}
\usepackage{dcolumn}
\usepackage{bm}
\usepackage{amsmath}
\usepackage{color}

\graphicspath{{./Figures/}}

\newcommand{\eqnref}[1]{Eq.~(\ref{eqn:#1})}

\newcommand{\secref}[1]{Sec.~\ref{sec:#1}}

\newcommand{\subsecref}[1]{Subsec.~\ref{subsec:#1}}

\newcommand{\figref}[1]{Fig.~\ref{fig:#1}}

\newcommand{\tableref}[1]{Table~\ref{table:#1}}

\def\slashchar#1{\setbox0=\hbox{$#1$}           
   \dimen0=\wd0                                 
   \setbox1=\hbox{/} \dimen1=\wd1               
   \ifdim\dimen0>\dimen1                        
      \rlap{\hbox to \dimen0{\hfil/\hfil}}      
      #1                                        
   \else                                        
      \rlap{\hbox to \dimen1{\hfil$#1$\hfil}}   
      /                                         
   \fi}

\begin{document}

\preprint{MITP/17-004}

\title{Collider constraints and new tests of color octet vectors}

\author{Malte Buschmann}
\email{buschmann@uni-mainz.de}
\affiliation{PRISMA Cluster of Excellence \& Mainz Institute for
  Theoretical Physics, Johannes Gutenberg University, 55099 Mainz,
  Germany}

\author{Felix Yu}
\email{yu001@uni-mainz.de}
\affiliation{PRISMA Cluster of Excellence \& Mainz Institute for
  Theoretical Physics, Johannes Gutenberg University, 55099 Mainz,
  Germany}


\begin{abstract}
We analyze the collider sensitivity for new colored resonances in $t
\bar{t}$, $b \bar{b}$, and $jj$ final states.  While searches in the
single production channel are model-dependent, the pair production
rate is model independent and the existing $(JJ)(JJ)$ and $4t$
searches impose strong constraints on the relevant branching
fractions, where $J = j$ or $b$.  We point out the missing,
complementary searches in the mixed decay modes, $t\bar{t}(jj)$,
$t\bar{t}(b\bar{b})$, and $(b\bar{b})(jj)$.  We propose analysis
strategies for the $t\bar{t}(jj)$ and $t\bar{t}(b\bar{b})$ decays and
find their sensivity surpasses that of existing searches when the
decay widths to tops and light jets are comparable.  If no other
decays are present, collective lower limits on the resonance mass can
be set at 1.5~TeV using 37~fb$^{-1}$ of 13~TeV data.
\end{abstract}


\maketitle
\tableofcontents

\section{Introduction}
\label{sec:Introduction}

Searches for new particles and new forces beyond the Standard Model
(BSM) are a critical endeavor for the ATLAS and CMS experiments at the
Large Hadron Collider (LHC), but such results have thus far come up
null.  The largest rates come from colored particle pair production,
where leading order cross sections can be calculated knowing only the
color charge and mass~\cite{Han:2010rf}.  Even so, the colored
particle decay patterns and corresponding collider signatures are
highly model dependent.

In models with an extended color gauge symmetry~\cite{Hill:1991at,
  Chivukula:1996yr, Frampton:1987dn, Martynov:2009en, Frampton:2009rk,
  Simmons:2011aa}, massive color octet vectors, such as
colorons~\cite{Bai:2010dj}, are heavy cousins of the familiar gluon
and can decay universally to Standard Model (SM) quark--anti-quark
pairs.  Axigluons~\cite{Hall:1985wz, Frampton:1987dn, Bagger:1987fz}
are similarly motivated by ascribing chiral projection operators to
the parent gauge groups, which necessitates new fermions transforming
non-trivially under the extended gauge group to cancel anomalies.
Similar phenomenology occurs in models of universal extra dimensions,
where the massive color octet vectors arise as Kaluza-Klein (KK)
excitations of the gluon~\cite{Appelquist:2000nn, Rizzo:2001sd,
  Macesanu:2002db}.  In Randall-Sundrum models~\cite{Randall:1999ee,
  Randall:1999vf} with SM fields propagating in the
bulk~\cite{Davoudiasl:1999tf, Grossman:1999ra, Pomarol:1999ad,
  Chang:1999nh, Huber:2000ie}, the KK gluons exhibit flavor-dependent
couplings to quark pairs, and preferentially decay to the heavy flavor
quarks as a result of the localization of the bulk fermion
wavefunction profile~\cite{Dicus:2000hm, Davoudiasl:2000wi,
  Lillie:2007yh, Casagrande:2008hr, Agashe:2013kxa}.

At the LHC, CMS and ATLAS have searched for color octet vector
resonances in the paired dijet channel~\cite{Chatrchyan:2013izb,
  Khachatryan:2014lpa, CMS:2016pkl, ATLAS:2012ds, ATLAS:2016sfd,
  ATLAS:2017gsy}, most recently constraining colorons with a 100\%
branching fraction to light jets to be heavier than
1.5~TeV~\cite{ATLAS:2017gsy}.  Searches for four
tops~\cite{CMS:2013xma, Sirunyan:2017tep, ATLAS:2012hpa, Aad:2015kqa,
  TheATLAScollaboration:2016gxs, ATLAS:2016gqb, ATLAS:2016btu} also
constrain the possible $t\bar{t}$ decays of pair-produced resonances.
Searches for TeV-scale dijet resonances~\cite{CMS:2017xrr,
  Aaboud:2017yvp} and $t\bar{t}$ resonances~\cite{Chatrchyan:2012cx,
  Chatrchyan:2012yca, Aad:2012dpa, Aad:2012ans, Aad:2013nca,
  Aad:2015fna, TheATLAScollaboration:2016wfb} offer complementary
probes compared to the pair-production searches, since such searches
scale with the individual production coupling~\cite{Dobrescu:2013coa}.

Given the possibility that the color octet vector has flavor dependent
branching fractions to quark pairs, and because color octet vectors
have a model independent pair production rate, the mixed decay
signature of a paired ditop and dijet resonance is strongly motivated.
This channel is complementary to the existing $(JJ)(JJ)$ and $4t$
searches and even offers superior sensitivity when the branching
fractions of the resonance to dijets and ditops are comparable.

In~\secref{theory}, we review the theory motivation and collider
phenomenology of massive color octet vectors.  In~\secref{collider},
we analyze the prospects for discovering color octet vector resonances
in the $t\bar{t}(jj)$ and $t\bar{t}(b\bar{b})$ mixed decay modes at
the LHC and compare with the current constraints.  We conclude
in~\secref{conclusion}.

\section{Theory background}
\label{sec:theory}

Massive color octet vector resonances arise in various different
beyond the Standard Model extensions, such as models with extended
color gauge groups or models with extra dimensions.  In the unbroken
phase of electroweak symmetry, the general interaction Lagrangian
between quarks and a massive color octet vector $X_\mu$ is
\begin{equation}
\mathcal{L} \supset g_s \left( 
\bar{Q}_L \gamma_\mu \lambda T^a X^{\mu \, a} Q_L + 
\bar{u}_R \gamma_\mu \kappa T^a X^{\mu \, a} u_R +
\bar{d}_R \gamma_\mu T^a \eta X^{\mu \, a} d_R \right) \ ,
\label{eqn:Lagrangian}
\end{equation}
where $g_S$ is the strong coupling constant, $T^a$ are the $SU(3)$
generators, and $\lambda$, $\kappa$, and $\eta$ are the
flavor-dependent couplings in the quark gauge basis.  While these
matrices must be symmetric by $CPT$, nonzero off-diagonal entries are
possible in principle and correspond to tree-level flavor violation.
Since these processes are strongly constrained by low-energy precision
flavor measurements, such as meson mixing measurements and $b \to X_s
\gamma$ transitions~\cite{Bona:2007vi, Chivukula:2013kw}, the simplest
ansatz is to adopt a flavor-universal coupling structure, evading the
most stringest flavor bounds.  In such a scenario, however, the
branching ratio for $X$ to light quarks vs.~tops is fixed, and there
is no model freedom in the complementarity between searching in dijet
vs.~ditop resonances.  Hence, we will focus on the scenario where the
$\lambda$, $\kappa$, and $\eta$ matrices are diagonal but not
universal in the quark gauge basis.  Moving to the quark mass basis
will then induce off-diagonal entries proportional to
Cabibbo-Kobayashi-Maskawa (CKM) mixing.  We will first briefly review
models of massive color octet vectors and then discuss the collider
and flavor physics phenomenology of the general
Lagrangian,~\eqnref{Lagrangian}.

\subsection{Color octet vectors from extended color gauge groups}
\label{subsec:extendedcolor}

In models with an extended color symmetry, such as a $SU(3)_1 \times
SU(3)_2$ gauge group, the two parent gauge groups are typically broken
to the diagonal subgroup by the vacuum expectation value (vev) of a
bifundamental, complex scalar field $\Sigma$~\cite{Chivukula:1996yr,
  Bai:2010dj}.  The diagonal subgroup is then identified with the SM
$SU(3)_c$ gauge group, which imposes the requirement
\begin{equation}
\frac{1}{g_s^2} = \frac{1}{h_1^2} + \frac{1}{h_2^2} \ ,
\end{equation}
where $h_1$ and $h_2$ are the gauge couplings of $SU(3)_1$ and
$SU(3)_2$, respectively.  The Goldstone modes of the complex scalar
field become the longitudinal degrees of freedom for the heavy color
octet vector, leaving one real and one pseudoreal color singlet scalar
and one real color octet scalar as the dynamical scalar fields below
$\langle \Sigma \rangle$.  Explicitly, the covariant derivative for
$\Sigma$ is
\begin{equation}
D^\mu \Sigma = (\partial^\mu - i h_1G_1^{\mu a}T^a + i h_2G_2^{\mu a}T^a) \Sigma \ ,
\end{equation}
and after Higgsing $SU(3)_1 \times SU(3)_2 \to SU(3)_c$, we get the SM
gluon and coloron fields,
\begin{eqnarray}
g^\mu = \cos\theta \, G_1^\mu + \sin\theta \, G_2^\mu \ , \notag \\
X^\mu = \sin\theta \, G_1^\mu - \cos\theta \, G_2^\mu \ ,
\end{eqnarray}
respectively, where $\theta = \tan^{-1} (h_1 / h_2)$ is the mixing
angle.

In this setup, different possibilities for the couplings
in~\eqnref{Lagrangian} originate by considering various charge
assignments of the quarks in the parent $SU(3)_1 \times SU(3)_2$ gauge
symmetry.  The main feature for universally coupled models is that all
flavor representations are assigned identically to the same gauge
representation, which ensures that the gauge symmetry commutes with
the global SM quark flavor symmetry.

In the coloron model~\cite{Chivukula:1996yr, Bai:2010dj}, all SM
quarks transform as $\square$ under one $SU(3)$ gauge group and
singlets under the other.  Hence, the coloron has flavor universal,
purely vector couplings to the SM quarks as
\begin{equation}
g_s \tan \theta \bar{q} \gamma^\mu T^a X_\mu^a q \ .
\label{eqn:coloroncoupling}
\end{equation}
An alternative prescription is the chiral color
model~\cite{Hall:1985wz, Frampton:1987dn, Frampton:1987ut}, where
left-handed (LH) and right-handed (RH) quark fields are charged under different
$SU(3)$ gauge groups.  This construction generally requires new
fermions to cancel anomalies, notably $SU(3)_1^2 \times U(1)_Y$ and
$SU(3)_2^2 \times U(1)_Y$, but this new matter content can be massive
and unobservable at colliders.  As a result, if the LH quarks
transform as $(\square, \mathbf{1})$ and the RH quarks
transform as $(\mathbf{1}, \square)$ under $SU(3)_1 \times SU(3)_2$, for
example, the resulting massive color vector interaction with quarks is
\begin{equation}
g_s \bar{q} \gamma^\mu T^a X_\mu^a (\tan \theta P_L - \cot \theta
P_R) q \ ,
\label{eqn:axigluoncoupling}
\end{equation}
where $X_\mu^a$ is commonly referred as an axigluon in the literature.

To motivate a non-universal yet diagonal coupling structure
in~\eqnref{Lagrangian}, we can straightforwardly assign different
quark flavors to different gauge representations under $SU(3)_1 \times
SU(3)_2$.  For example, the topcolor model charges the third
generation quarks differently than the first two generations, with
$Q_L^{1, 2} \sim (\square, \mathbf{1})$, $Q_L^3 \sim (\mathbf{1},
\square)$, $u_R^{1, 2} \sim (\square, \mathbf{1})$, $u_R^3 \sim
(\mathbf{1}, \square)$, $d_R^{1, 2, 3} \sim (\square,
\mathbf{1})$~\cite{Hill:1991at}.  This assignment also requires
additional matter to cancel anomalies, for which a minimal solution
involves two electroweak singlet quarks transforming as $(\mathbf{1},
\square)$ and $(\square, \mathbf{1})$, each with hypercharge $-2/3$.
The corresponding massive color octet vector does not have
flavor-changing couplings in the quark gauge basis, but instead
features distinct couplings to light and heavy generation quarks:
\begin{equation}
g_s \cot \theta (\bar{t} \gamma^\mu T^a X_\mu t + 
\bar{b}_L \gamma^\mu T^a X_\mu b_L) + g_s \tan \theta 
(\bar{b}_R \gamma^\mu T^a X_\mu b_R + 
\sum\limits_{i = 1 \dots 4} \bar{q}_i \gamma^\mu T^a X_\mu q_i) \ ,
\end{equation}
with $\tan \theta = h_1 / h_2$ as before.  Although topcolor models
are generally motivated by composite Higgs scenarios triggered by top
quark condensation~\cite{Hill:1991at}, we will only focus on the
motivated possibility that $X_\mu$ has non-universal couplings.

\subsection{Color octet vectors from extra dimensions}
\label{subsec:extradimensions}

Models with extra spacetime dimensions provide an alternative
framework for realizing massive color octet vector
resonances~\cite{Appelquist:2000nn, Rizzo:2001sd, Macesanu:2002db}.
In such models, the SM fields are the lowest-lying states of a
Kaluza-Klein tower, whose masses and dynamics result from solving the
five-dimensional equations of motion~\cite{Davoudiasl:1999tf,
  Grossman:1999ra, Pomarol:1999ad, Chang:1999nh, Huber:2000ie}.  In
minimal universal extra dimensions~\cite{Datta:2010us}, the level-2 KK
gluon obtains a coupling to SM quarks from one-loop diagrams with
level-1 KK particles running in the loop.  These couplings are
generated from boundary conditions on the KK gluon and the bulk
masses, which provide the only source of translational invariance
breaking, and read
\begin{align}
& g_s T^a \gamma^\mu \frac{1}{\sqrt{2}} \frac{1}{16 \pi^2} \log \left( \frac{\Lambda}{\mu} \right)^2 \left[ P_L (\frac{1}{8} g_1^2 + \frac{27}{8} g_2^2 - \frac{11}{2} g_s^2) + P_R (Y_{u, d} \ g_1^2 - \frac{11}{2} g_s^2) \right] \ ,
\label{eqn:mUEDcoupling}
\end{align}
where $Y_u = 2$ for up-type quarks, $Y_d = 1/2$ for down-type quarks,
$\Lambda$ is an ultraviolet scale larger than the inverse size of the
extra dimension, and $\mu$ is the renormalization scale to evaluate
the coupling.

In Randall-Sundrum warped scenarios~\cite{Randall:1999ee,
  Randall:1999vf}, the fermion mass hierarchy can be explained by
allowing fermions to propagate in the bulk, where the observed charged
fermion mass hierarchy originates as $\mathcal{O}(1)$ differences in
bulk mass parameters.  Typically, the KK mass scale must then be
$\mathcal{O}(5-10~\text{TeV})$ to satisfy low-energy flavor violation
probes, especially $\bar{K}-K$ mixing~\cite{Casagrande:2008hr}, but
this scale can be lowered in the case that the bulk fermions obey a
flavor symmetry~\cite{Chen:2009gy, Chen:2009hr}.  Since KK parity is
absent, the first KK gluon decays to SM zero-mode quarks, and this
coupling is given by calculating the wavefunction overlap between the
zero modes and the KK gluon in the extra dimension.  Using the general
coupling structure in~\eqnref{Lagrangian} and identifying $X_\mu$ with
the first KK gluon,
\begin{align}
\lambda^{ij} &\approx \dfrac{m_X}{\sqrt{2} M_{\text{KK}}} \left( \dfrac{1}{\sqrt{2L}} \delta_{ij} - \sqrt{2L} F(c_{Q_i}) F(c_{Q_j}) \right) \ ,
\end{align}
where RH up-type couplings are obtained by replacing $\lambda
\rightarrow \kappa$ and $Q \rightarrow u$, RH down-type couplings are
obtained by replacing $\lambda \rightarrow \eta$ and $Q \rightarrow
d$, $m_X \approx 2.4 M_{\text{KK}}$ is fixed by the boundary
conditions of the 5D gluon, $L$ is the length of the extra dimension,
and $c_Q$, $c_u$, and $c_d$ are the bulk mass parameters for LH and RH
quark fields~\cite{Agashe:2013kxa}.  In order to reproduce the known
SM quark masses, these bulk mass parameters must be chosen to maximize
the top quark wavefunction overlap with the TeV-scale infrared brane
while the wavefunctions for the light quarks are skewed towards the
ultraviolet brane.  As a result, the first KK gluon preferentially
decays to top quarks, with branching fractions that can exceed
80\%~\cite{Agashe:2013kxa}.

\subsection{Non-universal couplings and flavor constraints}
\label{subsec:flavor}

As we have emphasized, massive color octet vectors arise in numerous
models of beyond the Standard Model physics.  Their collider and
flavor physics phenomenology depends crucially on the particular
$\lambda$, $\kappa$, and $\eta$ structure defined
in~\eqnref{Lagrangian} that is realized in a given model.  Given the
stringent constraints on tree-level flavor changing neutral
currents~\cite{Bona:2007vi, Chivukula:2013kw}, we adopt
flavor-diagonal couplings for $\lambda$, $\kappa$, and $\eta$ in the
gauge basis.  Nevertheless, flavor violating effects are still induced
in interactions of LH down quarks by the rotation to the quark mass
basis.  From~\eqnref{Lagrangian}, we rotate the quark fields to the
mass basis by $V_u u_L = u_L^m$, $V_d d_L = d_L^m$, $U_u u_R = u_R^m$,
and $U_d d_R = d_R^m$, giving
\begin{align}
\mathcal{L} &\supset \bar{u}_L^m g_s t^a \slashed{X}^a V_u \lambda V_u^\dagger u_L^m
+ \bar{d}_L^m g_s t^a \slashed{X}^a V_d V_u^\dagger V_u \lambda V_u^\dagger V_u V_d^\dagger d_L^m \nonumber \\
&+ \bar{u}_R^m g_s t^a \slashed{X}^a U_u \kappa U_u^\dagger u_R^m
+ \bar{d}_R^m g_s t^a \slashed{X}^a U_d \eta U_d^\dagger d_R^m \ .
\end{align}
We see that the $\lambda$, $\kappa$, and $\eta$ matrices can be chosen
such that the effective interaction matrices $\tilde{\lambda}$,
$\tilde{\kappa}$, and $\tilde{\eta}$ are diagonal in the quark mass
basis,
\begin{align}
\tilde{\lambda} = V_u \lambda V_u^\dagger \ , \ 
\tilde{\kappa} = U_u \kappa U_u^\dagger \ , \ 
\tilde{\eta} = U_d \eta U_d^\dagger \ .
\end{align}
The corresponding LH down quark interactions have small, off-diagonal
entries induced by $V_{CKM} \equiv V_u
V_d^{\dagger}$~\cite{Gedalia:2010rj, Olive:2016xmw},
\begin{align}
\tilde{\lambda}_D \equiv V_d V_u^{\dagger} V_u \lambda V_u^\dagger V_u
V_d^{\dagger} = V_{CKM}^{\dagger} \tilde{\lambda} V_{CKM} \ ,
\end{align}
and hence $X_\mu$ mediates tree-level flavor-changing neutral currents
(FCNCs).  Since the strongest FCNC constraints come from
$\bar{K}_0-K_0$ mixing~\cite{Bona:2007vi}, we minimize the impact of
these constraints by assuming $\tilde{\lambda}_{11} =
\tilde{\lambda}_{22} \neq \tilde{\lambda}_{33}$, which leads to
$\tilde{\lambda}_{D, 12} \approx \tilde{\lambda}_{D, 21} \approx -(3.3
- 1.3i) \times 10^{-4} (\tilde{\lambda}_{33} - \tilde{\lambda}_{11})$,
using global fit values for the CKM elements~\cite{Olive:2016xmw}.  A
tree-level exchange of $X_\mu$ can be matched to the four-fermion
operator~\cite{Bona:2007vi, Casagrande:2008hr, Bai:2011ed,
  Haisch:2011up, Altmannshofer:2012ur, Chivukula:2013kw}
\begin{align}
O_K^1 = (\bar{d}_\alpha \gamma_\mu P_L s_\alpha) (\bar{d}_\beta
\gamma^\mu P_L s_\beta) \ ,
\end{align}
with the Wilson coefficient
\begin{align}
C_K^1 = \frac{-1}{6} \frac{\tilde{\lambda}_{D, 12}^2}{M_X^2} \ .
\end{align}
Using the requirement $\Lambda > 1.1 \times 10^3$~TeV from Re~$C_K^1$
and $\Lambda > 2.2 \times 10^4$~TeV from
Im~$C_K^1$~\cite{Bona:2007vi}, we can constrain the maximum size of
$\tilde{\lambda}_{11}$, $\tilde{\lambda}_{33}$, assuming the other is
vanishing:
\begin{align}
\text{Re}~C_K^1: &\quad M_X > 130~\text{GeV} g_s \text{ max}
(|\tilde{\lambda}_{11}|, |\tilde{\lambda}_{33}|) \\ 
\text{Im}~C_K^1: &\quad M_X > 2700~\text{GeV} g_s \text{ max}
(|\tilde{\lambda}_{11}|, |\tilde{\lambda}_{33}|) \ .
\end{align}
We see that these constraints are easily satisfied for
$\mathcal{O}(\text{TeV})$ scale color octets as long as
$\tilde{\lambda} \sim \mathcal{O}(0.1)$.  Moreover, with this
structure, $\tilde{\lambda}_{11} = \tilde{\lambda}_{22}$, the stronger
constraints come from $\bar{B}_d^0-B_d^0$ and $\bar{B}_s^0-B_s^0$
measurements, where $|C_{B_d}^1|$ requires $\Lambda > 1.0 \times
10^3$~TeV and $|C_{B_s}^1|$ requires $\Lambda > 240$~TeV.  Using the
tree-level, CKM-induced off-diagonal elements of $\tilde{\lambda}_D$,
we obtain
\begin{align}
|C_{B_d}^1|: &\quad M_X > 3500~\text{GeV} g_s \text{ max} (|\tilde{\lambda}_{11}|, |\tilde{\lambda}_{33}|) \\
|C_{B_s}^1|: &\quad M_X > 4000~\text{GeV} g_s \text{ max} (|\tilde{\lambda}_{11}|, |\tilde{\lambda}_{33}|) \ ,
\end{align}
which are again weakened to the sub-TeV scale for $\tilde{\lambda}
\sim \mathcal{O}(0.1)$.  As a result, we can realistically expect
$\mathcal{O}(\text{TeV})$ color octet vectors with dominant branching
fractions to either tops, $\tilde{\lambda}_{33} \gg
\tilde{\lambda}_{11}$ or jets, $\tilde{\lambda}_{11} \gg
\tilde{\lambda}_{33}$, entirely consistent with flavor bounds.  We
remark that an alternative assumption of the flavor structure can
relax these bounds further.  In particular, if instead
$V_{CKM}^\dagger \tilde{\lambda} V_{CKM}$ is diagonal, then the
tree-level flavor violating couplings are shifted to the LH up-quark
couplings~\cite{Bai:2011ed, Haisch:2011up}.  We can define
$\tilde{\lambda}' = V_{CKM}^\dagger \tilde{\lambda} V_{CKM}$ as
diagonal, and then the LH up-quark couples via $V_{CKM}
\tilde{\lambda}' V_{CKM}^\dagger$, which is constrained by
$D_0-\bar{D}_0$ mixing, where $\Lambda > 700$~TeV~\cite{Bona:2007vi}.
The corresponding bound is
\begin{align}
\text{Im}~C_D^1: &\quad M_X > 36~\text{GeV} g_s \text{ max}(|\tilde{\lambda}_{11}|,
|\tilde{\lambda}_{33}|) \ .
\end{align}
The $\Delta F = 1$ constraints, including $b \to X_s \gamma$ and $b
\to X_s g$ penguin amplitudes, are also immediately satisfied if the
$X^\mu$ couplings are aligned in the down-quark
sector~\cite{Chivukula:2013kw, Haisch:2011up}.  

\subsection{Collider phenomenology of color octet vectors}
\label{subsec:pheno}

Color octet vectors can be singly and doubly produced at colliders.
The single-production rate scales with the corresponding partial width
into $q \bar{q}$, $q = d, u, s, c$, since single production via gluons
is forbidden at tree-level.  In our scenario, the induced
flavor-violating couplings $\tilde{\lambda}_D$ are at most $5\% \times
\tilde{\lambda}_{11}, \tilde{\lambda}_{33}$, which we will neglect and
hence approximate $\tilde{\lambda}_D \approx \tilde{\lambda}$.  The
partial width of $X$ to a pair of quarks $q$ is
\begin{equation}
\Gamma (X \to \bar{q} q) = \dfrac{\alpha_s m_X}{12} 
\left( (g_L^2 + g_R^2)(1 - \dfrac{m_q^2}{m_X^2}) + 
6 \frac{m_q^2}{m_X^2} g_L g_R \right) 
\left( 1 - 4 \frac{m_q^2}{m_X^2} \right)^{1/2} \ ,
\label{eqn:Xtoqqwidth}
\end{equation}
where $g_L$ and $g_R$ are the corresponding diagonal entries in
$\tilde{\lambda}$, $\tilde{\kappa}$, and $\tilde{\eta}$\footnote{For
  reference, the generalized flavor violating partial width is
\begin{align}
\Gamma (X \to \bar{q}_i q_j, \bar{q}_j q_i) &= 
\dfrac{\alpha_s m_X}{6} 
\left( ((g_L^{ij})^2 + (g_R^{ij})^2) (1 - \frac{1}{2} (\frac{m_i^2}{m_X^2} + \frac{m_j^2}{m_X^2}) - \frac{1}{2} (\frac{m_i^2}{m_X^2} - \frac{m_j^2}{m_X^2})^2)
+ 6 \frac{m_i m_j}{m_X^2} g_L^{ij} g_R^{ij} \right) \nonumber \\
&\times \left( (1 - \frac{m_i^2}{m_X^2} - \frac{m_j^2}{m_X^2})^2 - 4 \frac{m_i^2 m_j^2}{m_X^4} \right)^{1/2} \ ,
\label{eqn:Xtoqiqjwidth}
\end{align}
where $g_L^{ij} = g_L^{ji}$ and $g_R^{ij} = g_R^{ji}$.}.  If the
masses of the quarks can be neglected compared to the $X$ mass, the
partial width simplifies to
\begin{equation}
\Gamma (X \to \bar{q} q) = \dfrac{\alpha_s (g_L^2 + g_R^2) m_X}{12} \ .
\end{equation}
The branching fractions for $jj$, $b\bar{b}$, and $t\bar{t}$ final
states are then simply ratios of the corresponding sum of squared
couplings.  We see that large branching fractions to top quarks can
easily be realized by increasing $\tilde{\kappa}_{33}$, while large
branching fractions to bottom quarks corresponds to increasing
$\tilde{\eta}_{33}$.  This is the effective description of the warped
extra dimension scenario with $t_R$ and $b_R$ wavefunction profiles
peaked close to the infrared brane~\cite{Agashe:2013kxa}, and small
hierarchies in these third generation couplings are consistent with
electroweak precision tests~\cite{Bai:2011ed, Haisch:2011up}.

As previously highlighted, the single dijet resonance constraints
scale with the overal partial width into light flavor quarks.
Pair-production rates, however, are robustly calculable knowing only
$m_X$ and its color octet representation.  Hence, all searches in
pair-production modes provide important complementary reach compared
to single resonance searches.  The current LHC analyses focus on the
simplest topologies, with $X X^* \to (JJ) (JJ)$, with $J = j$ or $b$
inclusively~\cite{Chatrchyan:2013izb, Khachatryan:2014lpa,
  CMS:2016pkl, ATLAS:2012ds, ATLAS:2016sfd, ATLAS:2017gsy} and $X X^*
\to 4t$~\cite{CMS:2013xma, Sirunyan:2017tep, ATLAS:2012hpa,
  Aad:2015kqa, TheATLAScollaboration:2016gxs, ATLAS:2016gqb,
  ATLAS:2016btu}.  In the case where the $X \to b\bar{b}$ decay width
is preferred, additional signal discrimination is easily gained by
requiring $b$-tags.  The orthogonal $(b\bar{b})(b\bar{b})$,
$(b\bar{b})(jj)$, and $(jj)(jj)$ signal regions would then have
enhanced and complementary sensitivity compared to the current
$(JJ)(JJ)$ searches.  If the coupling to tops is preferred, then the
$4t$ search and our proposed searches in the $t\bar{t}(jj)$ and
$t\bar{t}(b\bar{b})$ mixed decay channels are critical.  In
particular, the $t\bar{t}(jj)$ and $t\bar{t}(b\bar{b})$ mixed decay
searches offer substantial improvements in covering the sensitivity
gap when the $X$ decay widths to tops and light quarks are comparable.
In the next section, we describe our collider analyses optimized for
the $t\bar{t}(jj)$ and $t\bar{t}(b\bar{b})$ final states.


\section{Collider analyses of the mixed channels, $t\bar{t}(jj)$ and $t\bar{t}(b\bar{b})$}
\label{sec:collider}

We analyze pair-produced resonances in a new mixed decay mode, $p p
\to XX^* \to t\bar{t} (jj)$ and $p p \to XX^* \to t\bar{t}
(b\bar{b})$.  Although there are numerous possibilities for the top
decays, we mainly focus on the semi-leptonic and fully leptonic final
states, which provide clean handles for tagging the reconstructible
$t\bar{t}$ and dijet systems.  Of course, the main SM background is
the irreducible $t \bar{t} + $ jets background, where other single
boson and diboson + jets backgrounds are subleading after requiring
multiple $b$-jets.

We construct a \textsc{FeynRules} v2.3.26~\cite{Alloul:2013bka} model
using Universal FeynRules Output~\cite{Degrande:2011ua} to perform
leading order Monte Carlo event simulation with \textsc{MadGraph}~5
v2.4.3~\cite{Alwall:2014hca}, interfaced with
\textsc{Pythia}~v8.2~\cite{Sjostrand:2007gs, Sjostrand:2014zea} for
showering and hadronizations.  We remark that the leading order
calculation allows a direct comparison to existing $(JJ)(JJ)$ and $4t$
search limits~\cite{Chatrchyan:2013izb, Khachatryan:2014lpa,
  CMS:2016pkl, ATLAS:2012ds, ATLAS:2016sfd, ATLAS:2017gsy,
  CMS:2013xma, Sirunyan:2017tep, ATLAS:2012hpa, Aad:2015kqa,
  TheATLAScollaboration:2016gxs, ATLAS:2016gqb, ATLAS:2016btu}.  Since
the signal will rely on tagging two top candidates, we simulate the SM
background, $t\bar{t}j$ with up to two additional jets at leading
order, using \textsc{Sherpa} v.2.1.0~\cite{Gleisberg:2008ta}.  We
rescale the background by a flat $K$-factor of 1.5, adopted from
\textsc{Sherpa}+\textsc{BlackHat}~\cite{Gleisberg:2008ta,
  Berger:2008sj}.  The inclusive SM $t\bar{t}$ + jets production cross
section is known at next-to-next-to-leading order and
next-to-next-to-leading log (NNLL) precision~\cite{Czakon:2013goa},
whereas differential cross sections can be obtained at next-to-leading
order and NNLL precision~\cite{Guzzi:2014wia}.  Since our $X$
resonance should appear as a resonant peak over a smooth continuum
background, a $p_T$-dependent $K$-factor should not significantly
affect our projected results.

\subsection{Event selection strategy}
\label{subsec:selection}
We first discuss the semi-leptonic top decay channel, where our signal
process is $XX^* \to b \bar{b} \ell^\pm \nu jj (JJ)$, with $J = j$ or
$b$.  Jets are clustered using the anti-$k_T$
algorithm~\cite{Cacciari:2008gp} with $R = 0.5$, and we require jets
to have $p_T > 50$~GeV and $|\eta| < 4.9$.  We identify $b$-jets using
a \textsc{Delphes} v3.1.2~\cite{deFavereau:2013fsa} detector
simulation, with a $p_T$- and $\eta$-dependent tagging efficiency of
about $70\%$ for displaced tracks from $B$-mesons within $\Delta R <
0.3$ from the main jet axis; the charm misidentification rate is
roughly $15\%$ and the light flavor mistag rate is $0.1\%$.

Events must have at least 5 jets, at least two of which must be
$b$-tagged and one must be untagged.  The leading $b$-tagged and
untagged jets must have $p_T \geq 250$~GeV, and the subleading
untagged jet, if present, must have $p_T \geq 80$~GeV.  Events must
also have exactly one isolated lepton with $p_T > 20$~GeV and $|\eta|
< 2.5$, with isolation using \textsc{Delphes} default parameters.  We
furthermore cut on missing transverse energy (MET), requiring
$\slashed{E}_T > 80$~GeV, as well as signal mass-optimized $H_T =
\Sigma_j |p_{T, j}| \geq (4/3) \, m_X$.

Naively, the $X \to JJ$ signal resonance is extracted from forming the
invariant mass of the two leading jets.  The jet combinatorics,
however, present a major hurdle against using the dijet and ditop
masses to discriminate the signal from the irreducible background.
Thus, the main goal of our collider strategy is to solve the
combinatorial ambiguity, taking advantage of the resonant high mass
dijet signal, the large $p_T$ of the resonant dijet system, and the
relevant angular spread between the $b$-jets, lepton, and jets.
Difficulties in resolving multijet combinatorial ambiguities are
discussed in, {\it e.g.},~\cite{Rajaraman:2010hy}.

For this purpose, we define two signal regions.  We target the
$b\bar{b}$ mode by selecting events with more than two $b$-tagged
jets, while the $jj$ mode is optimized by requiring exactly two
$b$-tagged jets.  For the $b\bar{b}$ decay, we assume that the leading
$b$-tagged jet comes from the $X \to JJ$ decay directly, not from
$t\bar{t}$. We then reconstruct the $b\bar{b}$ system by adding the
remaining leading jet, whether tagged or untagged.  We find that this
reconstruction best reflects the dijet resonance in spite of the
underlying combinatorial ambiguity.

For the $jj$ mode, we start with the leading light jet $j_1$ and add
to it the hardest light jet $j_2$ which satisfies $\Delta R_{j_1 j_2}
\leq \pi$.  We also add in the next hardest light jet $j_3$ with
$\Delta R_{j_1 j_3}\leq\pi$ if $j_1$ and $j_2$ are not balanced in
$p_T$, when $p_{T, j_2} / p_{T, j_1} \leq 0.15$.  This follows the
hemisphere intuition, where the $X \to jj$ decays should be untagged
and relatively hard.  Adding the third hardest light jet accounts for
the wide-angle final state radiation of our signal quarks.

\begin{figure}[tbh!]
\includegraphics[width=0.48\textwidth]{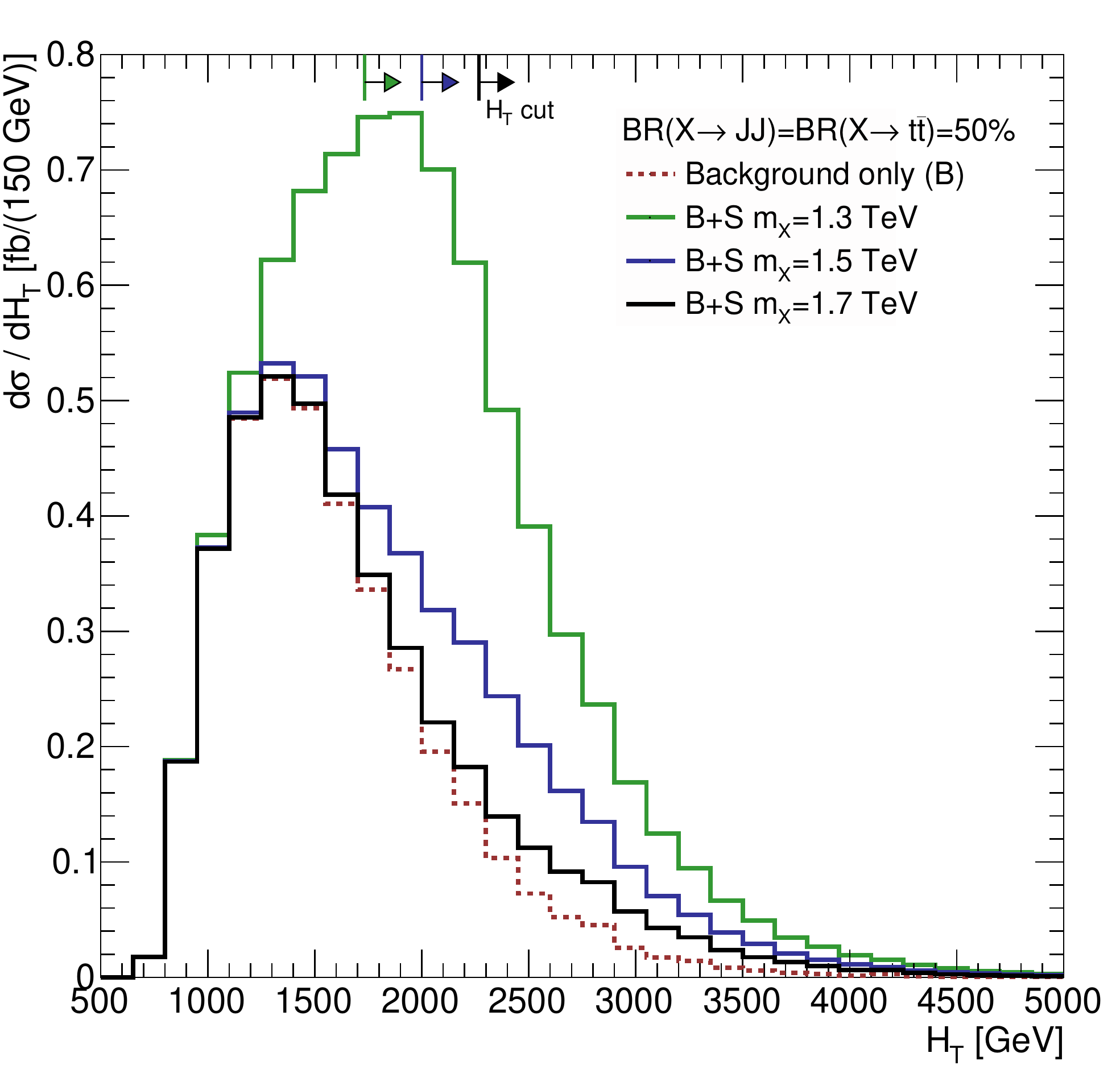}
\includegraphics[width=0.48\textwidth]{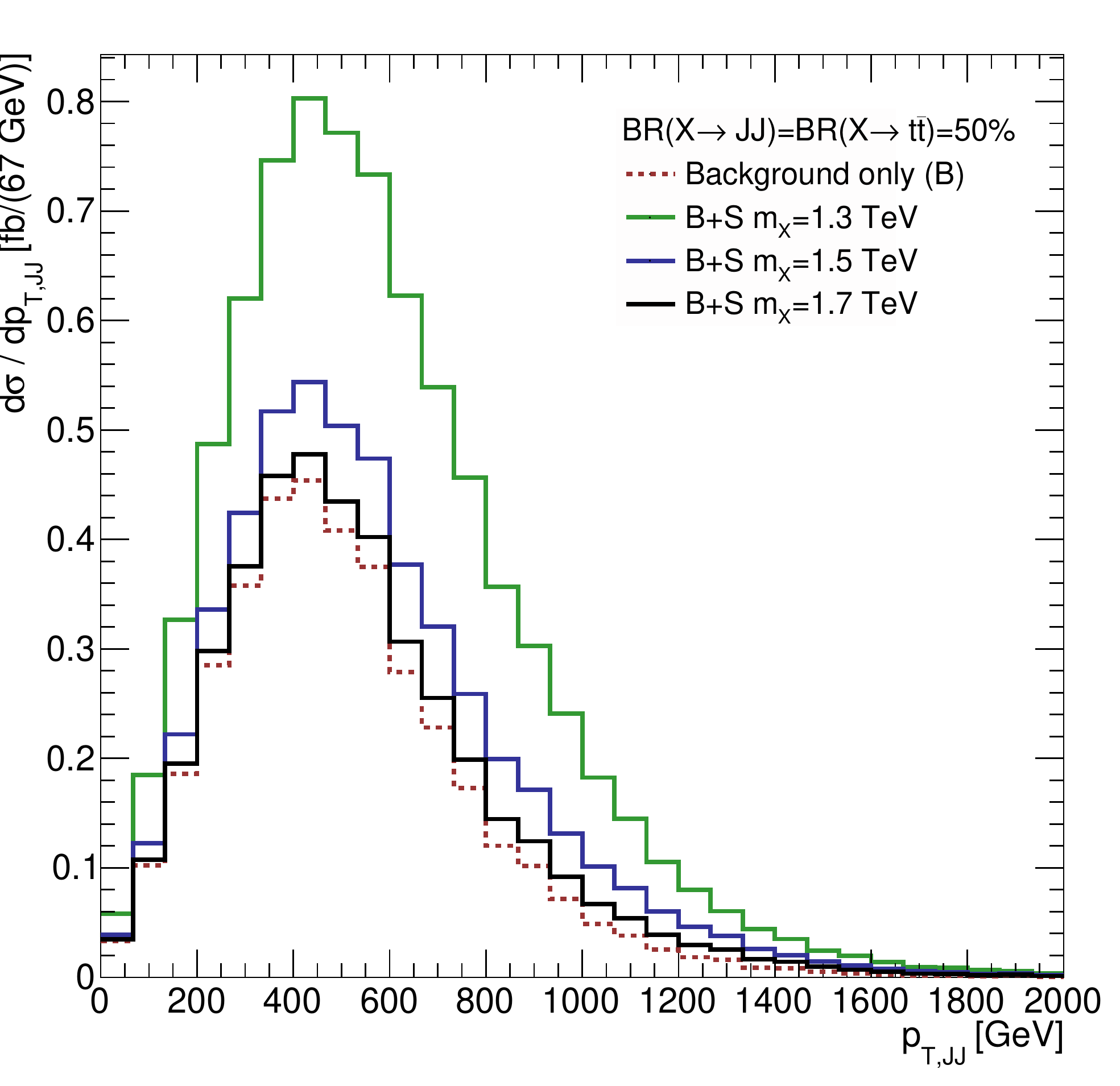}\\
\includegraphics[width=0.48\textwidth]{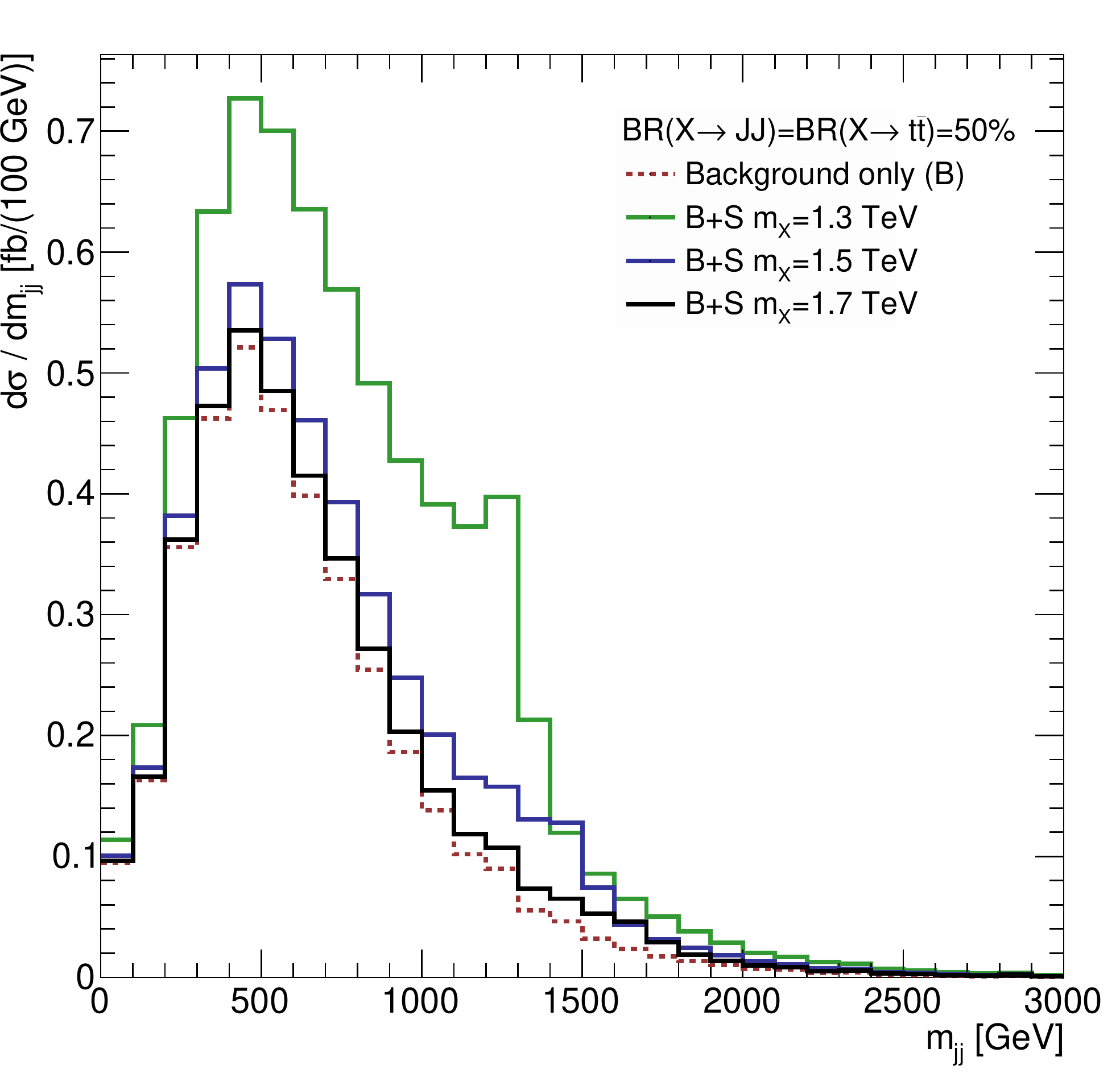}
\includegraphics[width=0.48\textwidth]{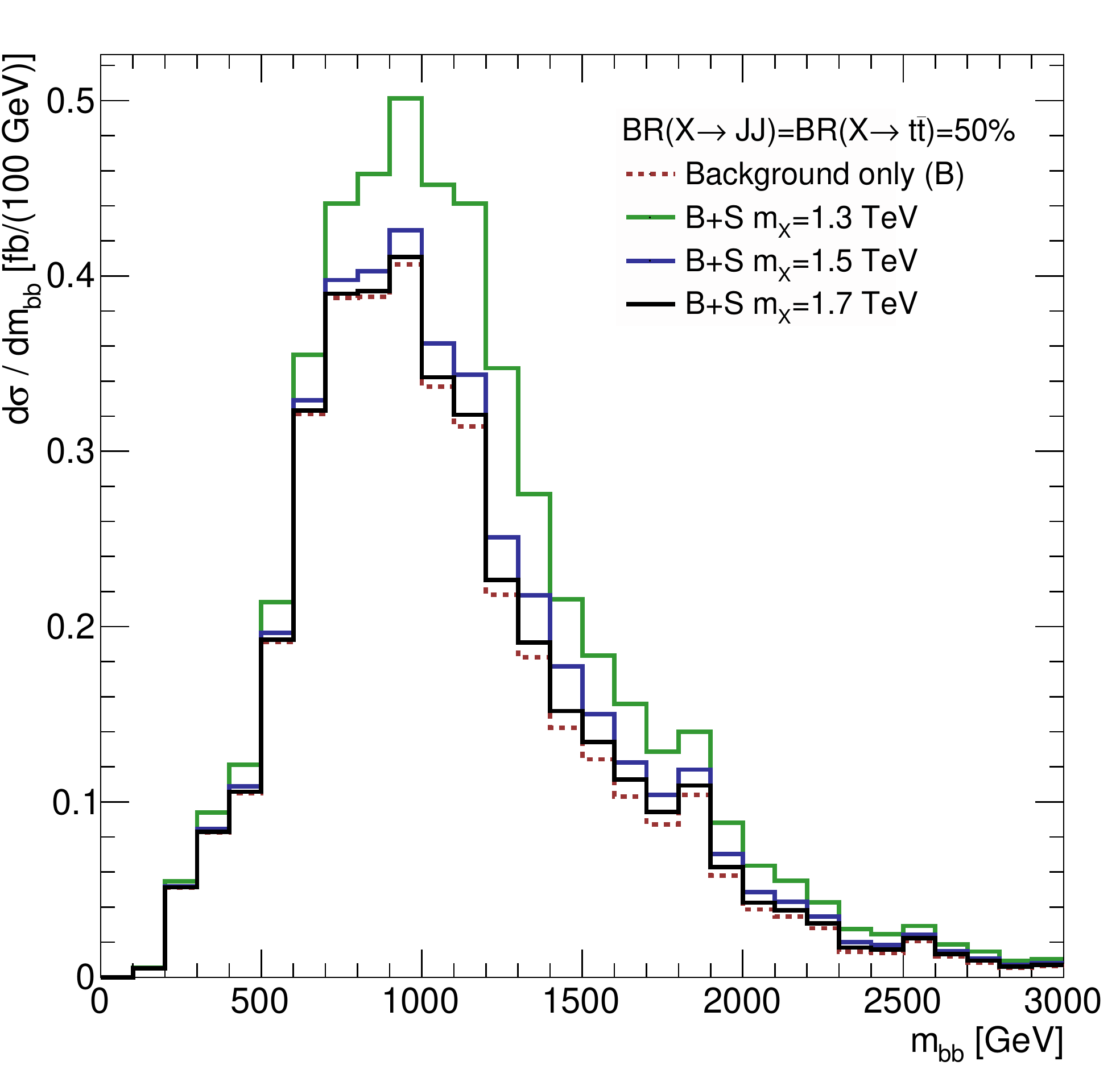}
\caption{Differential distributions in the semi-leptonic analysis for
  $H_T$ (upper left), $p_{T, JJ}$ (upper right), $m_{jj}$ (lower
  left), and $m_{bb}$ (lower right), where the invariant mass
  definitions are described in the main text.  We show distributions
  for the SM $t\bar{t} + $ jets background (red, dashed), stacked
  background + $m_X = 1.3$~TeV signal (green, solid), background +
  $m_X = 1.5$~TeV signal (blue, solid), background + $m_X = 1.7$~TeV
  signal (black, solid).  Differential distributions are presented
  after jet selection and $\slashed{E}_T$ cuts but before the $H_T$ to
  show the unsculpted dijet invariant mass spectra.}
\label{fig:kinematics}
\end{figure}

For both $b\bar{b}$ and $jj$ decay modes, we construct the dijet
invariant mass as our final kinematic discriminant.  The signal and
background cut efficiencies for the semileptonic analysis are shown
in~\tableref{semilep_cutflow}, while the most salient kinematic
distributions for the background stacked with different signal
hypotheses are shown in~\figref{kinematics}.  The upper left panel
in~\figref{kinematics} shows the hardening of the $H_T$ distribution
coming from the signal jets.  The dijet $p_{T, JJ}$ distribution,
however, only offers an overal rate shift from the additional signal
events and no strong correlation with the signal $X$ mass, which is
because of the combinatorial ambiguity among the jets.  The bottom
panels show the invariant mass distributions in the $jj$ and
$b\bar{b}$ targetted modes.  Again, the broad peak structure arises
mostly from the combinatorial ambiguity in capturing the correct
signal jets to reconstruct the resonance.  We note that $b$-tagging
efficiency and the combinatorial ambiguity also cause $X \to jj$
events to populate the $m_{bb}$ signal region and vice versa.

\begin{center}
\begin{table}[tb!]
  \begin{tabular}{| l | c c c | c c c |}
    \hline
     {\bf Semi-Leptonic Search}& \multicolumn{3}{c|}{Signal} & \multicolumn{3}{c|}{Background $t\bar{t}$+jets}\\
     Signal mass $m_X$ 		& 1.3 TeV & 1.5 TeV 	& 1.7 TeV	& 1.3 TeV 	& 1.5 TeV	& 1.7 TeV\\\hline
     Event selection [fb] ($N_J \geq 5$ with $N_j \geq 1$, $N_b \geq 2$; 	&&&&&&\\
     $N_\ell=1$; $p_{T,j}^\text{leading},\ p_{T,b}^\text{leading}>250$ GeV; $p_{T,\ell}>20$ GeV)	& 1.52  	& 0.45  	& 0.13 	& 5.8  	& 5.8 	& 5.8\\
     $\slashed{E}_T > 80$~GeV  & 84\% 	& 87\% 	& 89\%	& 64\% 	& 64\%	& 64\%\\
     $H_T \geq (4/3)\,m_X$ 		& 83\% 	& 79\% 	& 76\%	& 32\% 	& 19\%	& 10\%\\
     Remaining cross section [fb]	& 1.06 	& 0.31  	& 0.09 	& 1.23  	& 0.71 	& 0.39 \\
    \hline
  \end{tabular}
\caption{Cut flow for different resonance masses $m_X$ and dominant
  background $t\bar{t}$+jets for the semi-leptonic search. All
  branching ratios are applied to signal and background when quoting
  cross sections. We normalize the signal assuming $\text{Br}(X \to
  t\bar{t})=\text{ Br}(X \to JJ)=50\%$. }
    \label{table:semilep_cutflow}
  \end{table}
\end{center}

The search in the fully leptonic top quark decay channel is very
similar.  We again select jets and leptons as in the semi-leptonic
channel, but we require only 4 jets, at least two of which are
$b$-tagged, and also require exactly two isolated leptons of opposite
charge.  We loosen the MET cut, $\slashed{E}_T > 50$~GeV, and keep the
$H_T$ cut, $H_T \geq (4/3) \, m_X$.  To avoid the anticipated $Z+$jets
background we veto events with $m_{\ell\ell}\leq 115$~GeV.  Again we
distinguish between $b\bar{b}$ and $jj$ dijet resonance decay modes,
using the same method as the semi-leptonic search.  The corresponding
cut flow is presented in~\tableref{fullylep_cutflow}.  The final
signal rates lead to the same relative signal to background ratios as
the semi-leptonic analysis, but the absolute rate is only an $\approx
5\%$ additional contribution compared to the semi-leptonic channel.

\begin{center}
\begin{table}[tbh!]
  \begin{tabular}{| l | c c c | c c c |}
    \hline
     {\bf Fully Leptonic Search} & \multicolumn{3}{c|}{Signal} & \multicolumn{3}{c|}{Background $t\bar{t}$+jets}\\
     Signal mass $m_X$ 		& 1.3 TeV & 1.5 TeV 	& 1.7 TeV	& 1.3 TeV 	& 1.5 TeV	& 1.7 TeV\\\hline
     Event selection [fb] ($N_J \geq 4$ with $N_b \geq 2$; 	&&&&&&\\
     $N_\ell = 2$; $p_{T,j}^\text{leading},\ p_{T,b}^\text{leading}>250$ GeV; $p_{T,\ell}>20$ GeV)	
     						& 0.148  	& 0.045  	& 0.009 	& 0.75  	& 0.75 	& 0.75\\
     $\slashed{E}_T > 50$~GeV  & 96\% 	& 96\% 	& 97\%	& 91\% 	& 91\%	& 91\%\\
     $m_{\ell\ell}\geq 115$~GeV  & 66\%	& 68\%	& 70\%	& 37\%	& 37 \%	& 37\%\\
     $H_T \geq (4/3)\,m_X$ 	& 63\% 	& 57\% 	& 48\%	& 21\% 	& 11\%	& 2.7\%\\
     Remaining cross section [fb]	& 0.059 	& 0.017  	& 0.003 	& 0.053  	& 0.027 	& 0.007 \\
    \hline
  \end{tabular}
\caption{Cut flow for a different resonance masses  $m_X$ and dominant
  background $t\bar{t}$+jets for the fully leptonic search. All branching
  ratios are applied to signal and background when quoting cross
  sections. We normalize the signal assuming $\text{Br}(X \to
  t\bar{t})=\text{ Br}(X \to JJ)=50\%$. }
    \label{table:fullylep_cutflow}
  \end{table}
\end{center}

\subsection{Comparison with current searches}
\label{subsec:comparison}

We now compare the projected sensitivity from the mixed $t\bar{t}
(JJ)$ searches in combination with the recasted exclusions from ATLAS
and CMS for $(JJ)(JJ)$ and $4t$ searches.  We also show the single
production limits for dijet and ditop resonance searches.  Since we
assume that our new physics state $X$ only decays to quark pairs, we
can express the pair production constraints in the branching fraction
vs.~mass plane, where
\begin{equation}
\text{Br}(X \to t\bar{t}) = 1 - \text{ Br}(X \to JJ) \ .
\end{equation}
So, the pair production limits can be translated according to:
\begin{equation}
\left. \sigma_{\text{excl}} (JJ)(JJ) \right|_{m_X} = \left. \sigma (pp \to
XX^*)\right|_{m_X} \times \left( \text{Br}_{JJ} \right)^2 \ ,
\end{equation}
with $\text{Br}_{JJ} = \text{Br}(X \to JJ)$.  The $4t$ constraint and
the $t\bar{t} (JJ)$ projected exclusion after replacing
$(\text{Br}_{JJ})^2$ with $(\text{Br}_{tt})^2$ and $2(\text{Br}_{JJ}
\ \text{Br}_{tt})$, respectively, where $\text{Br}_{tt} = \text{ Br}(X
\to t\bar{t})$.

\begin{figure}[tbh!]
\includegraphics[width=0.95\textwidth]{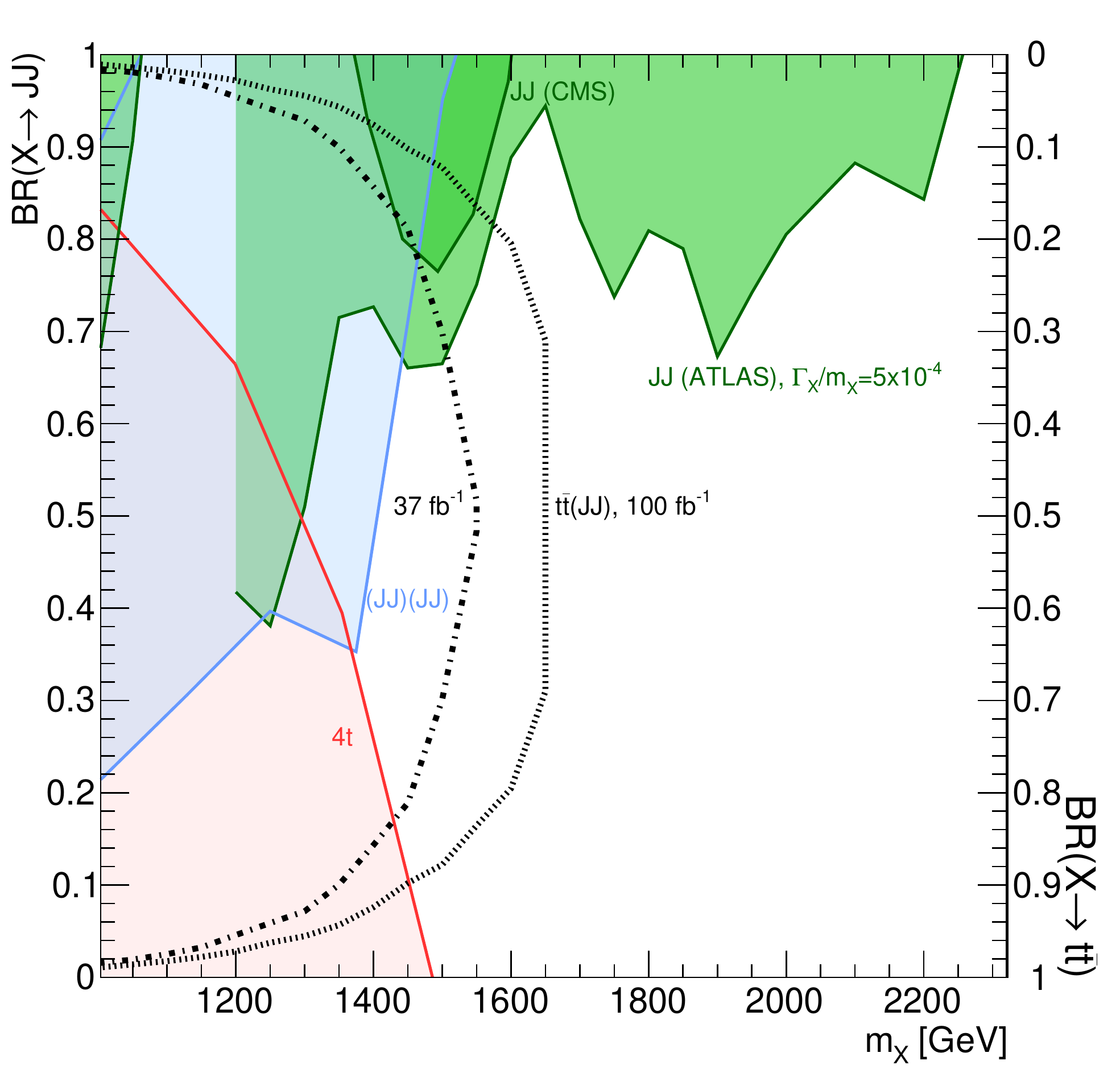}
\caption{Exclusion limits for different resonance masses as a function
  of $\text{ Br}(X \to JJ) = 1-\text{ Br}(X \to t\bar{t})$.  We show
  our limit in black for 37 fb$^{-1}$ (dot-dashed) and 100 fb$^{-1}$
  (dashed) of integrated luminosity.  We also show current limits from
  $4t$ (red)~\cite{ATLAS:2016btu}, $(JJ)(JJ)$
  (blue)~\cite{ATLAS:2017gsy}, and $(JJ)$ (green)~\cite{CMS:2017xrr,
    Aaboud:2017yvp} searches.}
    \label{fig:exclusion}
\end{figure}

The single dijet resonance limits are determined only after specifying
the total width of the resonance.  If the total width is narrow, a
reference cross section can be rescaled by the partial width into
light quarks.  We see that
\begin{align}
\left. \sigma_{\text{excl}} (pp \to (JJ)_{\text{res}}) \right|_{m_X} &= 
\sigma_{\text{width}} \text{ Br}_{JJ} A 
= \sigma_{\text{ref}} \text{ Br}_{JJ}^2 \frac{\Gamma_{\text{tot, width}}}{\Gamma_{qq, \text{ ref}}} A 
\end{align}
In the case of $t\bar{t}$ resonances, the above constraint becomes a
bounded requirement on the branching fraction to tops.  Hence, the
$t\bar{t}$ resonance constraint follows
\begin{align}
\left. \sigma_{\text{excl}} (pp \to (t\bar{t})_{\text{res}}) \right|_{m_X} &\geq 
\sigma_{\text{width}} \text{ Br}_{tt}
= \sigma_{\text{ref}} (1 - \text{ Br}_{tt}) \text{ Br}_{tt} \frac{\Gamma_{\text{tot, width}}}{\Gamma_{qq, \text{ref}}} \ .
\label{eqn:Brttbound}
\end{align}
Note that here we drop the acceptance factor since the experiments
unfold the acceptance when presenting their results.  We combine all
four orthogonal signal regions of the semi-leptonic and fully leptonic
search and compute the $95\%$ C.L.~limits based on the respective $JJ$
invariant mass shapes, where we assume a 10\% systematic uncertainty
on signal and background.  The result can be found
in~\figref{exclusion}.

\begin{figure}[tbh!]
\includegraphics[width=0.45\textwidth]{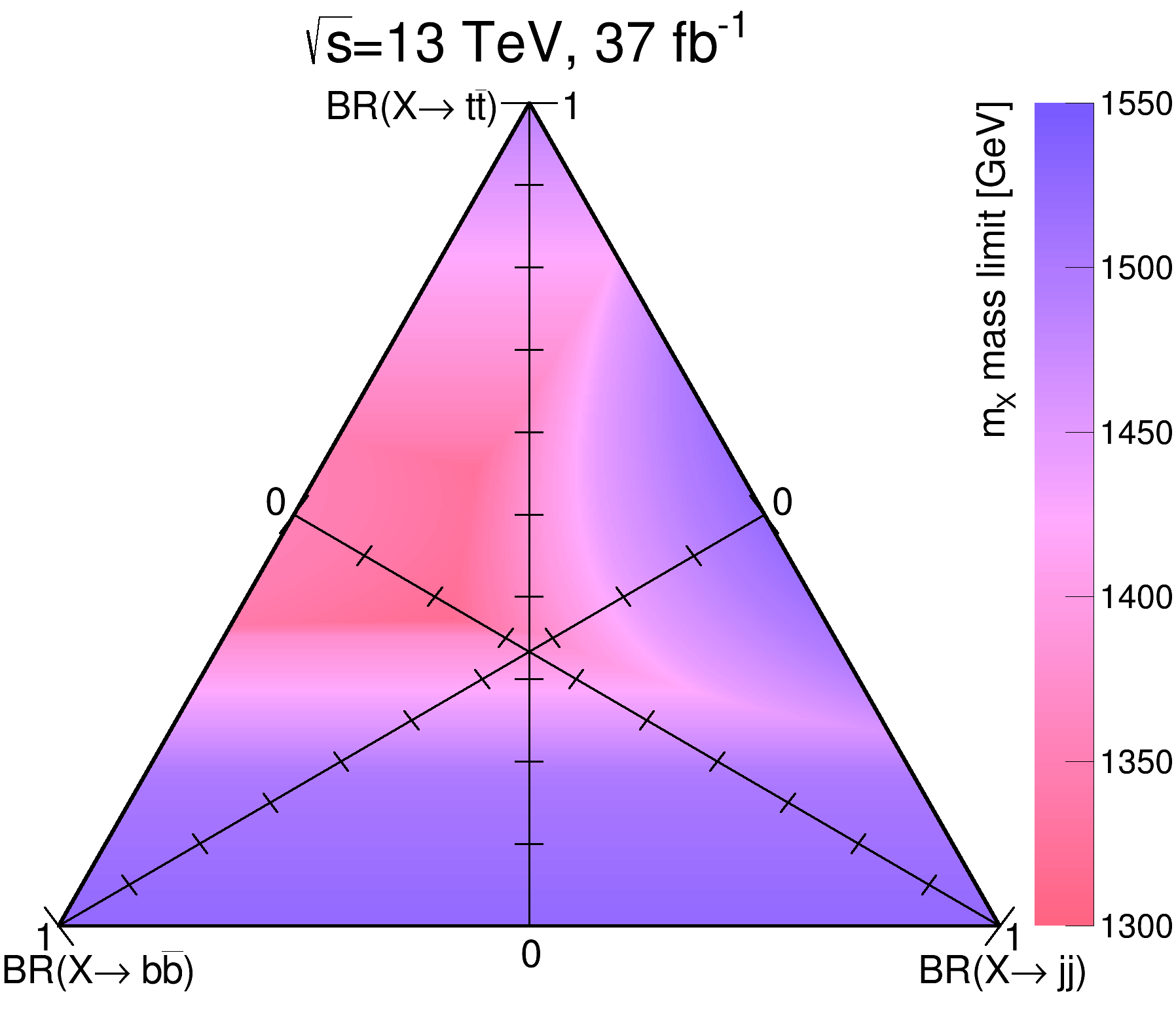}
\includegraphics[width=0.45\textwidth]{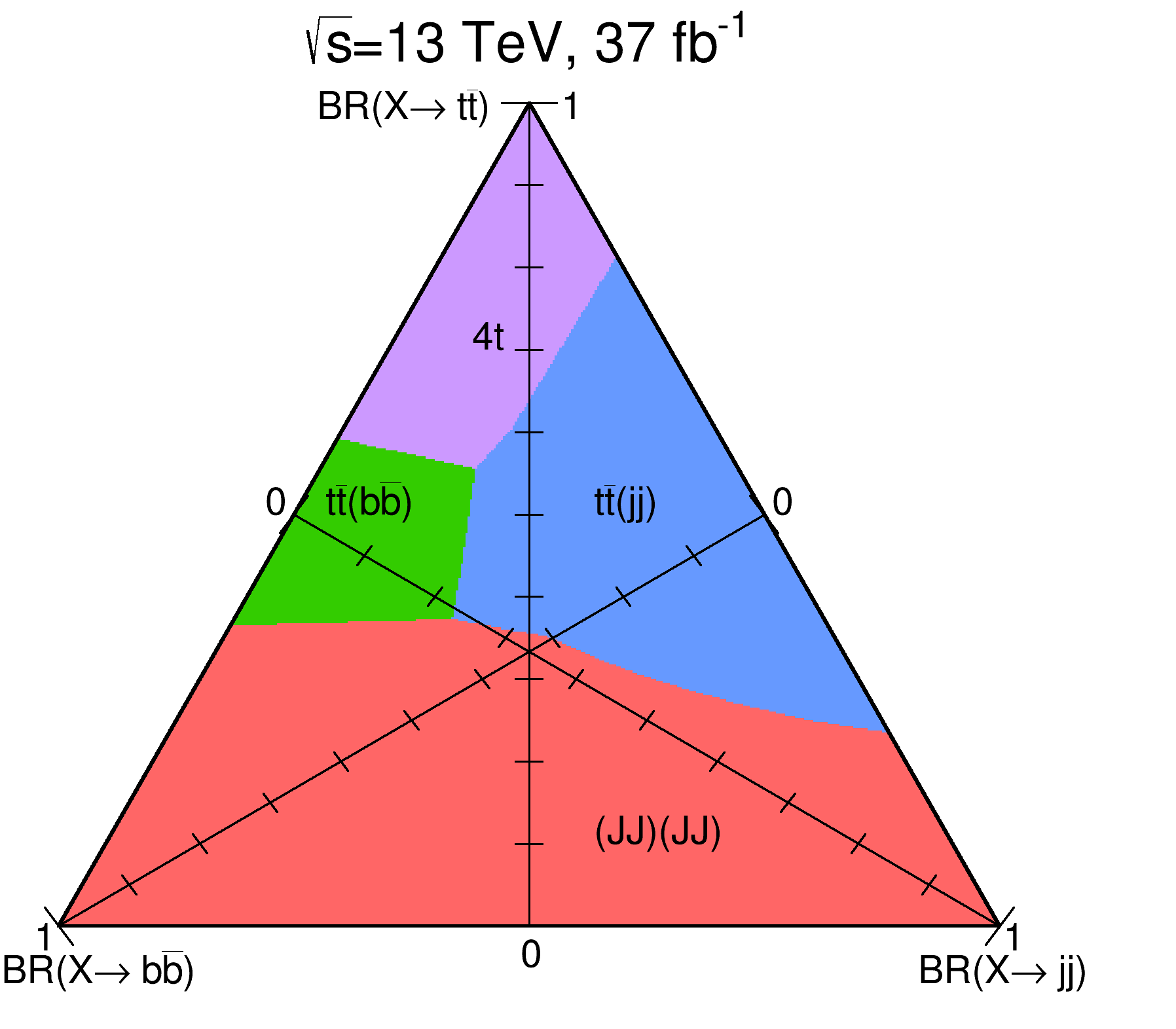}
\caption{Left: Strongest mass limits when the various $t\bar{t}$,
  $b\bar{b}$, and $jj$ decay channels are compared.  Right: Search
  region that gives the best sensitivity.}
    \label{fig:reach}
\end{figure} 

We remark that the width of our resonance, $\Gamma_X = 5 \times
10^{-4} m_X$, corresponds to diagonal entries of $\tilde{\lambda}$,
$\tilde{\kappa}$, $\tilde{\eta} \lesssim 0.1$.  As discussed
in~\subsecref{flavor}, such LH quark couplings to $X$ readily satisfy
flavor violation bounds from meson oscillation measurements and
neutral current transitions given $m_X > 1$~TeV.  On the other hand,
these couplings generally arise from mixing the SM quarks with heavy
vectorlike states, since otherwise perturbativity in the parent
extended color gauge symmetry is violated~\cite{Dobrescu:2013coa}.
The corresponding flavor violation bounds, collider constraints,
electroweak precision observable tests are more model-dependent, but
realistic and complete scenarios can be constructed~\cite{Bai:2011ed,
  Haisch:2011up, Chivukula:2013kw}.

As mentioned previously, the sensitivity from single production
compared to the sensitivity from pair production strongly depends on
the ratio $\Gamma_X/m_X$.  Any direct resonance bounds from $JJ$
searches could in principle be evaded completely with a suitably small
choice of $\Gamma_X$, but choosing $\Gamma_X / m_X = 5 \times 10^{-4}$
allows complementary dijet constraints from CMS and
ATLAS~\cite{CMS:2017xrr, Aaboud:2017yvp}.  We remark that the
$t\bar{t}$ resonance limits~\cite{TheATLAScollaboration:2016wfb} are
absent from~\figref{exclusion} for $\Gamma_X / m_X = 5 \times
10^{-4}$.  These constraints are only relevant once $\Gamma_X /
m_X \gtrsim 7 \times 10^{-4}$ for $m_X$ around 1~TeV.  As shown
in~\eqnref{Brttbound}, the $t\bar{t}$ resonance limit is symmetric
around $\text{Br}_{tt} = 50\%$ since the maximum rate in this channel
corresponds to equal partial widths to dijets and ditops.

The strongest existing bounds in~\figref{exclusion} are therefore from
$(JJ)(JJ)$~\cite{ATLAS:2017gsy} and $4t$~\cite{ATLAS:2016btu}
searches, which are clearly optimal for their respective
$\text{Br}_{JJ}$ and $\text{Br}_{tt}$ corners.  The mass reach of both
searches weakens by about 250~GeV in the intermediate regime, however,
leaving significant room for our dedicated $t\bar{t}(JJ)$ search to
explore.

In~\figref{exclusion}, we assume that all quark couplings are flavour
universal except for the top.  We relax this assumption
in~\figref{reach}, allowing both branching fractions $\text{Br}(X \to
b\bar{b}) = \text{ Br}_{bb}$ and $\text{Br}_{tt}$ to float.  The
results are presented in an equilateral triangle since the sum of the
$jj$, $b\bar{b}$ and $t\bar{t}$ branching fractions must equal 100\%
in our model.  The shading in the left panel of~\figref{reach}
shows the lower limit on the resonance mass $m_X$ as a function of the
three branching fractions. The right panel of~\figref{reach}
indicates which particular dedicated search yields the corresponding
lower mass limit.  We see that our new search channels, $t\bar{t}(jj)$
and $t\bar{t}(b\bar{b})$, overtake the sensitivity in the central
areas of the triangle compared to the existing $(JJ)(JJ)$ and $4t$
searches.

We reiterate that a complete characterization of $t\bar{t}$,
$b\bar{b}$, and $jj$ decay channels for pair production colored
resonances would necessitate optimizing the current $(JJ)(JJ)$ search
into $(b\bar{b})(b\bar{b})$, $(b\bar{b})(jj)$, and $(jj)(jj)$ signal
regions.  In particular, the $(b\bar{b})(b\bar{b})$ search would bear
striking similarities with the current searches for pair production of
the 125~GeV Higgs~\cite{Aaboud:2016xco, CMS:2016foy}, where the main
novelty would be varying the $(b\bar{b})$ mass window to test for new
resonances.  The multijet background, however, is very challenging to
simulate and thus substantial statistics in the $b$-tagged backgrounds
would be required to suitably smooth paired invariant mass spectrum in
the $(b\bar{b})(b\bar{b})$ and $(b\bar{b})(jj)$ channels.

\section{Conclusion}
\label{sec:conclusion}

In this work, we have highlighted the $t\bar{t}(jj)$ and
$t\bar{t}(b\bar{b})$ mixed decay channels of a massive, color octet
vector as new targets for ATLAS and CMS searches.  Hierarchies in the
underlying couplings of the $X$ resonance to light quarks, bottom
pairs, or top pairs are entirely consistent with low energy FCNC
constraints if the $X$ mass is above $1$~TeV and its couplings to
quarks are at most $0.1$.  As a result, the LHC provides the leading
reach to TeV-scale color octet vectors with variable couplings to
heavy and light flavor quarks by virtue of the model independent pair
production rate.

In our $t\bar{t}(JJ)$ analyses, we focused on resolving the jet
combinatorial ambiguity to reconstruct the dijet or dibottom
resonance.  In principle, reconstructing the $(t\bar{t})$ resonance is
also possible, but our scenario with its many resolved jets did not
afford any additional signal discrimination in this regard.

Nevertheless, our results show that new $t\bar{t}(jj)$ and
$t\bar{t}(b\bar{b})$ searches will fill a sensitivity gap between the
$(JJ)(JJ)$ and $4t$ searches.  This gap is clear in the
$\text{Br}_{JJ}$ vs.~$m_X$ plane, which itself provides a useful tool
for easily presenting the results from different collider searches of
pair-produced resonances.  We stress that a post-discovery scenario of
a new resonance greatly benefits from this complementary information,
where single and pair production modes combined with different decay
channels provide direct information about underlying Lagrangian
couplings.

\section*{Acknowledgments}
\label{sec:acknowledgments}

FY would like to thank R.~Sekhar Chivukula and Elizabeth Simmons for
helpful discussions during the 2015 Les Houches summer session, and
also Susanne Westhoff at the Mainz Institute for Theoretical Physics,
``The TeV Scale: A Threshold to New Physics?'' 2017 scientific
program.  FY would also like to acknowledge the hospitality of the
theory group at Fermi National Accelerator Laboratory while this work
was completed.  This research is supported by the Cluster of
Excellence Precision Physics, Fundamental Interactions and Structure
of Matter (PRISMA-EXC 1098).  The work of MB is moreover supported by
the German Research Foundation (DFG) in the framework of the Research
Unit ``New Physics at the Large Hadron Collider'' (FOR 2239).

\bibliography{referencelist}

\end{document}